\documentstyle[12pt]{article}
\def\pslash{p\kern-.5em\slash}
\def\kslash{k\kern-.5em\slash}
\def\beq{\begin{equation}}
\def\eeq{\end{equation}}
\def\beqa{\begin{eqnarray}}
\def\eeqa{\end{eqnarray}}
\def\MeV{\nobreak\,\mbox{MeV}}

\def\cbar{\overline{c}}

\def\bra#1{\langle #1|}
\def\ket#1{| #1\rangle}
\def\nucbra{\bra{ N}}
\def\nucket{\ket{ N}}

\begin{document}


\thispagestyle{empty}
\setcounter{page}{0}
\vspace*{48pt}
\begin{center}
{\large{\bf ON THE INTRINSIC CHARM COMPONENT OF THE NUCLEON}}
\end{center}
\vskip7mm
\begin{center}
F. S. NAVARRA, M. NIELSEN, C. A. A. NUNES and M. TEIXEIRA \\
\vskip3mm
Nuclear Theory and Elementary Particle Phenomenology Group\\
Instituto de F\'\i sica da Universidade de S\~ao Paulo,\\
Caixa Postal 20516, 01498-970 S\~ao Paulo, Brazil\\
\vskip5mm
\end{center}
\vskip15mm
\baselineskip=26pt plus1pt minus1pt

\vspace{2cm}

\noindent {\bf Abstract:}Using a $\overline D$ meson cloud model
we calculate the squared charm radius of the nucleon .
The ratio between this
squared radius and the ordinary baryon squared radius is
identified with the probability of ``seeing'' the intrinsic charm component
of the nucleon. Our estimate is compatible with those used to successfully
describe the charm production phenomenology.
\vfill
\eject

\noindent
\vskip7mm

In the early eighties there was a hope to understand charm production
solely in terms of perturbative QCD. Inspite of all uncertainties
in defining the scale , it would be in any case of the order of a few
GeV and therefore the coupling constant would be smaller than one.
As more and more data became available it became clear that perturbative
QCD alone was not enough to properly account for the  measured
differential cross sections. Higher order corrections did not make the
situation any better. The main problem was that the produced charmed particles
were too fast. In other words, there was a remarkable excess of particles
with large Feynman $x$ ( $x_F$ ).

Already at that time, the idea was advanced \cite{b},
that the  hadron wave function
contains a charm component even before undergoing a collision. This
component is originated in higher twist QCD interactions
inside the hadron. The so called ``intrinsic'' charmed pairs produced
by these interactions have nothing to do with usual sea quark pairs.
The crucial difference between them is that the intrinsic
charm is part of the valence system and therefore very fast
in contrast to the sea charm, which is slow. During the last years, an
intrinsic charm component was added to the perturbative QCD component in a
quantitative and systematic way \cite{vg} .
As a result , a very good description of
data was achieved. In order to obtain such good agreement
with experimental data the crucial point was the normalization of the
intrinsic charm component , $\sigma_{ic}$ of the hadron $+$ nucleon
$ \rightarrow c-\overline c$ X cross section .
The quantity $\sigma_{ic}$
is related to the probability of observing the intrinsic charm
component of the hadron, $P_{ic}$ \cite{vgb}.
It is very difficult (if not impossible)
to calculate this quantity from first principles. It was estimated from
a phenomenological analysis to be less than $1\%$ \cite{hm}.
In fact, $P_{ic}=0.3\%$ seems
to be the best value to describe recent data on charm production \cite{vg}.

A very important question is, of course, whether this $1\%$ of intrinsic
charm can be supported by any model based calculation. In ref. \cite{dg},
such a calculation was done using the MIT bag model. It was found that the
probability of finding a five-quark component $\ket{uudc\cbar}$
configuration bound within the nucleon bag is of 1 or $2 \%$,
in good agreement with the above mentioned phenomenological estimate.

In this note we  calculate $P_{ic}$ using an approach, which is completely
different and independent from that used in ref.~{[1-5]} and can therefore
be used as a cross-check to those estimates.

The existence of intrinsic charm is here associated with
low momentum components  of a virtual c-$\overline c$ pair in the nucleon.
At low momentum scales , the virtual pair lives a sufficiently long time to
permit the formation of charm hadronic components
of the nucleon wave function. The same argument is true for the strange
matrix elements.

Generally speaking, we can say that the proton is a fluctuating object, being
sometimes a neutron plus a pion, sometimes a strange hyperon plus kaon and
so on. It can be any combination of virtual hadrons possessing the right
quantum numbers. In particular, if charmed pairs pre-exist inside the nucleon,
it can oscilate into a charmed hyperon plus a $D$ meson, as e.g.,
by the process
\beq
p\rightarrow \Lambda_c + \overline D \rightarrow p \; .
\eeq

We calculate the intrinsic charm contribution to the matrix element
$\nucbra \cbar\gamma_\mu c\nucket$ arising from this virtual $\overline D$
meson cloud. The idea that intrinsic quark contributions to nucleon
matrix elements can be given by meson clouds is not new. It was used
in refs.\cite{mu,khp,kha,cfn} to estimate the intrinsic strangeness content
of the nucleon and it was suggested in ref \cite{b} as a picture to
understand the existence of intrinsic charm in the nucleon.

As in ref.\cite{mu}, we compute the $\overline D$ meson loops using an
effective meson-nucleon vertex characterized by a monopole form factor
\beq
F(k^2)=\frac{m^2-\Lambda^2}{ k^2-\Lambda^2} \; ,
\label{ff}
\eeq
and we introduce ``seagull'' terms in order to satisfy the Ward-Takahashi
(WT) identity. In eq. (2) $m$ is the meson mass and $\Lambda$ is the
effective cut-off. The inclusion of the meson-nucleon form-factors is
important to properly take into account the underlying nucleon structure
and its spatial extension. As shown in ref.\cite{kha}, when the sub-structure
of the nucleon is considered, it is the size of the proton, rather than
the masses involved in the loop, that determines the effective momentum
cut-off. We expect therefore the effective cut-off in the $\overline D$
meson-nucleon form factor to be approximately the same used in
the pion-nucleon or kaon-nucleon form factors.

The pseudoscalar meson-baryon coupling  for extended hadrons is
schematically given by
\beq
{\cal L}_{BBM} =-ig_{BBM} \bar{\Psi}\gamma_5 \Psi F(-\partial^2)
\phi \; ,
\label{la}
\eeq
where $\Psi$ and $\phi$ are baryon and meson fields respectively,
$F(k^2)$ is the form factor at the meson-baryon vertices and $k$
is the  momentum of the meson. The fact that the
nucleon-$\overline D$-$\Lambda_c$ coupling constant is not known is
not important here because we are mostly interested in arriving at some
upper limit to the intrinsic charm content of the nucleon and not
at definitive numerical predictions. Accordingly we will use
the pion-nucleon coupling constant as an upper limit to the
nucleon-$\overline D$-$\Lambda_c$ coupling constant.

We employ pointlike couplings between the current and the intermediate
meson and baryon. For the vector current one has
\beq
\bra{\Lambda_c(p^\prime)}\cbar\gamma_\mu c\ket{\Lambda_c(p)}=\overline
U(p^\prime) \gamma_\mu U(p)
\eeq
and
\beq
\bra{\overline D(p^\prime)}\cbar\gamma_\mu c\ket
{\overline D(p)}=-(p+p^\prime)_\mu
\eeq
in a convention where the $c$-quark has charm charge=+1.

The effective lagrangian eq.(\ref{la}) is non-local and this induces an
electromagnetic vertex current if the photon is present. In order to
maintain gauge
invariance we have to take into account the ``seagull vertex''
\beq
i\Gamma_\mu(k,q)=\pm g_{N\Lambda_c\overline D }
\gamma_5(q\pm 2k)_\mu
\frac{F(k^2) - F((q\pm k)^2)}{(q\pm k)^2 - k^2} \; ,
\label{seagull}
\eeq
which is generated via minimal substitution \cite{ohta}.
The upper and lower signs in eq.(6)
correspond to an incoming or outgoing meson respectively.

The three distinct contributions to the intrinsic form factors, associated
with processes in which the current couples to the baryon line (B)
(figure 1a) ,
to the meson line (M) (figure 1b) or to the meson-baryon vertex (V)
(figure 1c and 1d) in the loop are given by
\beq
\Gamma^B_\mu(p^\prime,p) = -ig^2_{N\Lambda_c\overline D}
\int \frac{d^4k}
{(2\pi)^4} \Delta(k^2) F(k^2) \gamma_5 S(p^\prime,k)
\gamma_\mu S(p,k) \gamma_5 F(k^2) \; ,
\label{bv}
\eeq
\beq
\Gamma^M_\mu(p^\prime,p) = ig^2_{N\Lambda_c\overline D}  \int
\frac{d^4k}{(2\pi)^4} \Delta((k+q)^2) (2k+q)_\mu \Delta(k^2)
F((k+q)^2) \gamma_5 S(p,k)  \gamma_5 F(k^2) \; ,
\label{mv}
\eeq
\beqa
\Gamma^V_\mu(p^\prime,p)& =& ig^2_{N\Lambda_c\overline D}  \int \frac{d^4k}
{(2\pi)^4} F(k^2) \Delta(k^2) \left[\frac{ (q+2k)_\mu}{ (q+k)^2-k^2}
\left(F(k^2)\, - F((k+q)^2)\right) \times \right.
\nonumber\\*[7.2pt]
& &\left.  \gamma_5 S(p-k) \gamma_5 - \frac{ (q-2k)_\mu}{
(q-k)^2-k^2} \left(F(k^2)-F((k-q)^2)\right) \gamma_5 S(p^\prime-k)
\gamma_5\right] \; .
\label{vv}
\eeqa

In the above equations
\beq
\Delta(k^2) = \frac{1}{k^2-m^2+i\epsilon}
\eeq
is the meson propagator and
\beq
 S(p,k) = \frac{1}{ \pslash-\kslash-M_\Lambda+i\epsilon}
\eeq
is the $\Lambda_c$ propagator and $p^\prime=p+q$ with $q$ being
the photon momentum. In figure 1 we show all momentum definitions.

With these amplitudes it is easy to show that the Ward-Takahashi
identity
\beq
q^\mu(\Gamma^B_\mu(p^\prime,p) + \Gamma^M_\mu(p^\prime,p) +
\Gamma^V_\mu(p^\prime,p)) = Q_c( \Sigma(p) - \Sigma(p^\prime)) \; ,
\label{WT}
\eeq
is satisfied. In eq.(12) $ Q_c $ is the nucleon charm charge,
$Q_c = 0$,
and $\Sigma(p)$ is the self-energy of the nucleon related to
the $\overline D~ \Lambda_c$ loop. The sum of the three amplitudes also
ensures the charge non-renormalization (or the Ward Identity)
\beq
(\Gamma^B_\mu + \Gamma^M_\mu + \Gamma^V_\mu)_{q=0} = Q_c\left(
-\frac {\partial} {\partial p^\mu} \Sigma(p)\right) = 0 \; .
\label{W}
\eeq

The intrinsic charm form factors are obtained by writing
these amplitudes in terms of the Dirac and Pauli form factors
\beq
\Gamma_\mu(p^\prime,p) = \gamma_\mu F_{1}^c(q^2) + i
\frac{\sigma_{\mu\nu}q^\nu} {2M_N} F_{2}^c(q^2) \; .
\eeq

The intrinsic squared charm radius of the nucleon is defined as
\beq
r^2_c  = \left.6\frac{\partial G_E^c(q^2)}{\partial q^2}
\right|_{q^2=0} \; ,
\label{r2}
\eeq
where $G_E^c(q^2)$ is the electric form factor introduced by Sachs \cite{sa}
\beq
G_E^c(q^2)=F_1^c(q^2) + \frac{ q^2}{4M_N^2}F_2^c(q^2)\; .
\eeq

The numerical results for $| r^2_c |$ are
shown in figure 2 , as a function of
the form factor cut-off $\Lambda$. The value of the coupling
and masses used are $M_N=939\MeV$, $M_{\Lambda_c}=2285\MeV$,
$m_{\overline D}=1865\MeV$ and $g_{N\Lambda_c \overline D}/\sqrt{4\pi}=
g_{N\pi N}/\sqrt{4\pi}=-3.795$

The intensity of a given proton fluctuation is associated with its average
squared radius $| r^2 |$. The larger is $| r^2 |$ , the more frequently we will
find the proton in that particular oscilation and the larger will be the
probability of ``seeing '' it.

We shall assume that the average barionic radius of the proton
$(r_p=[< r^2_B >]^\frac{1}{2}, \sim\, 0.72 fm )$ associated with the isoscalar
part of the electromagnetic current
is a good measure of the proton ``total size'', i.e., the size
which takes into account all possible fluctuations that couple to isoscalar
currents.
The intrinsic charm probability is then given by
$$
P_{ic} = \frac{| r^2_c |}{r^2_p} = 0.9 \% \eqno (13)
$$
where $| r^2_c | = 0.0047 fm^2$ is the average charm squared radius
calculated above with a cut-off $\Lambda = 1.2 $ GeV.
$P_{ic}$ is the ratio between the charm ``area'' and the
total proton ``area''.

We want to compare our results with those obtained by Donoghue and Golowich
in ref.~{[5]} for
the five quark components of the proton wave function ,
$ |uuds\overline s> $ and $ |uudq\overline q> $ ,
where q represents a light quark. We repeat then the
calculations for kaon and pion loops ( with the same cut-off $\Lambda$)
, obtaining the average strange
radius $| r^2_s | = 0.025 fm^2 $ and the average light quark radius
$| r^2_q | = 0.130 fm^2 $. Dividing these radii by the barionic squared
radius used above we obtain the probabilities $P_{is} = 5$ \%
and $P_{iq} = 25$ \% . The calculations done in ref.~{[5]} arrive at
$P_{is} = 16 $ \% and $P_{iq} = 31 $ \% . The discrepancy in the strange
sector suggests that the vector meson dominance model contribution
coming from the $\omega-\phi$ mixing ( see ref.~{[9]}) is really
important . In fact, it will change the result from $P_{is} = 5 $ \% to
$P_{is} = 10 $ \% [9]. As there is no experimental evidence for a
$\omega-J/\Psi$ mixing , the vector meson model will not contribute in the
charm sector. With the inclusion of the $\omega-\phi$ mixing our results
agree with those obtained in ref.~{[5]} within 6 \%.

The charm squared radius increases with $\Lambda$
(as it can be seen in figure 2) reaching $| r^2_c | = 0.016 fm^2$
at asymptotically large values of $\Lambda$. In this limit we would have
$P_{ic} = 3.0$ \%. Considering that we are overestimating the coupling
constant in the charm loop, this number can be taken as an upper limit for
the intrinsic charm probability in the context of our calculation scheme.
Our result seems to corroborate the previous estimates [1-5] .

In this work
we have only considered loops involving the particular combination
$\overline D$-$\Lambda_c$ . In principle we could include loops with
$\overline D$-$\Sigma_c$ and also with vector mesons. However in the case of
the intrinsic strangeness these  contributions were shown \cite{mu}
to be much less
important than the $\overline D$-$\Lambda_c$ loops. The $\Sigma_s$
couples very weakly to the nucleons and the vector mesons have large masses
, their contribution being thus supressed. In our case, due to the lack of
knowledge of the relevant couplings and cut-offs , no attempt is made
to go beyond the $\overline D$-$\Lambda_c$ loop. We expect this contribution
to be the most significant, specially in view of the very large values of the
coupling constant and cut-off used here. This might be sufficient for an
estimate of the order of magnitude of $P_{ic}$.
\vskip10mm
\noindent
\eject
{\bf AKNOWLEDGEMENTS}

\vskip7mm
This work was partially supported by FAPESP-Brazil and CNPq-Brazil. It
is a pleasure to thank R. Vogt for fruitful discussions.
\vfill
\eject
\eject

\eject
\noindent
{\Large\bf Figure Captions}\\ \\ \\
{\bf Figure 1.} Diagrams which contribute to the calculation of the
vertex function. Solid external lines represent the proton and solid
internal lines represent the  $\Lambda_c$. Dashed and wavy lines
represent the $\overline D$ and the vector current respectively.
\\
{\bf Figure 2.} The intrinsic charm mean square radius of the
nucleon as a function of the cut-off $\Lambda$ in the baryon-meson
form factor.
\eject

\end{document}